\documentclass[conference]{IEEEtran}
\IEEEoverridecommandlockouts
\usepackage{cite}
\usepackage{amsmath,amssymb,amsfonts}
\usepackage[linesnumbered, ruled, vlined]{algorithm2e}
\usepackage{algorithmic}
\usepackage{graphicx}
\usepackage{listings}
\usepackage{textcomp}
\usepackage{xcolor}

\makeatletter
\@ifundefined{showcaptionsetup}{}{
 \PassOptionsToPackage{caption=false}{subfig}}
\usepackage{subfig}
\makeatother

\usepackage{eso-pic}
\newcommand\AtPageUpperMyright[1]{\AtPageUpperLeft{
 \put(\LenToUnit{0.5\paperwidth},\LenToUnit{-1cm}){
     \parbox{0.5\textwidth}{\raggedleft\fontsize{9}{11}\selectfont #1}}
 }}
\newcommand{\conf}[1]{
\AddToShipoutPictureBG*{
\AtPageUpperMyright{#1}
}
}

\usepackage{tikz}
\usepackage{textcomp}
\usepackage{hyperref}
\newcommand\copyrighttext{%
  \footnotesize \textcopyright 2019 IEEE.  Personal use of this material is permitted.  Permission from IEEE must be obtained for all other uses, in any current or future media, including reprinting/republishing this material for advertising or promotional purposes, creating new collective works, for resale or redistribution to servers or lists, or reuse of any copyrighted component of this work in other works.
  DOI: \href{https://doi.org/10.1109/HPEC.2019.8916473}{10.1109/HPEC.2019.8916473}}
\newcommand\copyrightnotice{%
\begin{tikzpicture}[remember picture,overlay]
\node[anchor=south,yshift=10pt] at (current page.south) {\fbox{\parbox{\dimexpr\textwidth-\fboxsep-\fboxrule\relax}{\copyrighttext}}};
\end{tikzpicture}%
}

\definecolor{codegreen}{rgb}{0,0.45,0}
\definecolor{codegray}{rgb}{0.5,0.5,0.5}
\definecolor{codepurple}{rgb}{0.58,0,0.82}
\definecolor{backcolour}{rgb}{0.99,0.99,0.97}

\lstdefinestyle{mystyle}{
  backgroundcolor=\color{backcolour},   commentstyle=\color{codegreen},
  keywordstyle=\color{magenta},
  numberstyle=\tiny\color{codegray},
  stringstyle=\color{codepurple},
  basicstyle=\ttfamily\tiny,
  breakatwhitespace=false,         
  breaklines=true,                 
  captionpos=b,                    
  keepspaces=true,                 
  numbers=left,                    
  numbersep=5pt,                  
  showspaces=false,                
  showstringspaces=false,
  showtabs=false,                  
  tabsize=2
}

\def\BibTeX{{\rm B\kern-.05em{\sc i\kern-.025em b}\kern-.08em
    T\kern-.1667em\lower.7ex\hbox{E}\kern-.125emX}}
\begin{document}
\bstctlcite{IEEEexample:BSTcontrol}

\title{Exploration of Fine-Grained Parallelism for Load Balancing Eager K-truss on GPU and CPU\\
}

\author{
    \IEEEauthorblockN{Mark Blanco\IEEEauthorrefmark{1}, Tze Meng Low\IEEEauthorrefmark{2}
    \IEEEauthorblockA{\IEEEauthorrefmark{1}\IEEEauthorrefmark{2}
    \textit{Dept. of Electrical and Computer Engineering} \\
    \textit{Carnegie Mellon University}\\
    Pittsburgh, United States \\
    \{markb1 lowt\}@cmu.edu\\ }
    }
    \and
    \IEEEauthorblockN{Kyungjoo Kim\IEEEauthorrefmark{3}}
    \IEEEauthorblockA{\IEEEauthorrefmark{1}\IEEEauthorrefmark{3}
    \textit{Center for Computing Research} \\
    \textit{Sandia National Laboratories}\\
    Albuquerque, United States \\
    \{kyukim\}@sandia.gov}
}

\maketitle
\conf{2019 IEEE High Performance Extreme Computing Conference (HPEC)}
\copyrightnotice

\begin{abstract}
In this work we present a performance exploration on Eager K-truss, a linear-algebraic formulation of the K-truss graph algorithm. We address performance issues related to load imbalance of parallel tasks in symmetric, triangular graphs by presenting a fine-grained parallel approach to executing the support computation. This approach also increases available parallelism, making it amenable to GPU execution. We demonstrate our fine-grained parallel approach using implementations in Kokkos and evaluate them on an Intel Skylake CPU and an Nvidia Tesla V100 GPU. Overall, we observe between a 1.26-1.48x improvement on the CPU and a 9.97-16.92x improvement on the GPU due to our fine-grained parallel formulation. 
\end{abstract}


\section{Introduction}

The K-trusses of a graph $\mathcal G$ are highly connected subgraphs of $\mathcal G$
where each edge in a subgraph is an edge in at least $K-2$ distinct triangles in the 
subgraph~\cite{cohen2008trusses}.

In this work, we present a fine-grained parallel implementation of the Eager K-truss algorithm\cite{low2018linear}, a linear algebraic formulation of an edge-centric K-truss algorithm, on both the CPU and the GPU. The key observation is that the
existing Eager K-truss algorithm uses a coarse-grained approach to parallelism by dividing the edges into blocks that are distributed between parallel workers based on the common vertex that the edges are connected to; typically the `source' vertex. As each edge may be connected to different number of neighboring edges (i.e. edges that share the same vertex), the performance of this approach suffers from potential load imbalance. In addition, the Eager K-truss algorithm computes with an upper-triangular adjacency matrix, which means this load imbalance may be skewed significantly as the algorithm proceeds. 

We resolve the load imbalance problem by introducing a finer-grained task unit upon which parallel workers compute edge membership in triangles - the edge support values. This, in essence, introduces parallelism within each block of edges assigned to a processing element.

In Sections~\ref{sec:eager} and \ref{sec:fine-grained-eager}, we present details of our proposed algorithm. For an efficient implementation, we use a performance portable parallel programming model, Kokkos\cite{Edwards:2014:kokkos}. In particular, we report our K-truss performance for several fixed K values on NVIDIA V100 and Intel Skylake architectures. We report our results in millions of edges processed per second for each graph, and provide a summary table with timings in milliseconds for specific configurations.




%
\section{Background}
\label{sec:eager}

To keep this paper self-contained, we provide a brief description of the Eager K-truss algorithm in this section. For the detailed algorithmic derivation, we refer to Low et al.~\cite{low2018linear}.

\subsection{Linear Algebraic K-truss Algorithm}

Similar to many other K-truss implementations, Eager K-truss takes a two-step approach. Specifically, the first step computes the support of all edges, and the second step prunes edges whose support are below a specified threshold. This two-step approach is then repeated on the pruned graph(s) until no more edges can be removed. 
\begin{algorithm}[tb]
\small
\KwIn{$A$ is the adjacency matrix of input graph}
\KwOut{$S$ represents the support of edges in $A$}
\BlankLine                                                    
$converge \leftarrow false$\\
\While{not $converge$} {
  $S \leftarrow A^T A \circ A$ \tcp*[h]{Step 1: computeSupports}\\
  $M \leftarrow S \geq (k-2)$ \tcp*[h]{Step 2: pruneEdges}\\
  $A \leftarrow A \circ M$ \\
  $converge \leftarrow isUnchanged(M)$\\
  }
  \Return{$S$} \\
  \caption{Linear algebraic K-truss algorithm. Lowercase letters are vectors and capital letters are matrices.}
  \label{alg:linear-alg-ktruss}
\end{algorithm}

These two steps of the K-truss algorithm can generally be expressed using linear algebraic notation as described in Algorithm~\ref{alg:linear-alg-ktruss}, 
where $A$ is given as the adjacency matrix of the input graph, and $M$ is a binary matrix where $M[i,j]=1$ when $S[i,j] \geq (k-2)$ \cite{burkhardt_graphing_2017}. The $\circ$ operator represents an element-wise multiplication of the input operands.
Note that Step 1 computes matrix $S$, where each value at $S[i,j]$ is the number of triangles containing the edge between nodes $i$ and $j$.

\subsection{Eager K-truss Algorithm}
The Eager K-truss algorithm derives its name for the eager manner in which it updates the support values of all edges for each triangle that has been identified. Specifically, the support of all three edges that form a triangle is updated by the edge with the smallest two vertex labels.
%
By re-arranging $S \leftarrow A^T A \circ A$, using a block-partitioned matrix form, the first step of the Eager K-truss algorithm is described in Algorithm~\ref{alg:eager-ktruss}. To be consistent with the notation used in \cite{low2018linear}, we use upper-case, lower-case and Greek letters to represent matrices, column vectors, and scalar elements, respectively. Since Eager K-Truss specifically focuses on undirected and unweighted graphs, it uses the upper triangular adjacency matrix in its computations. Therefore, the adjacency matrix A used in algorithm \cite{low2018linear} is upper-triangular.

Iterating over the rows of the adjacency matrix $A$ and exploiting the symmetry of the undirected graph, the computation of the support matrix $S$ follows two updating rules: lines 3 and 4 depicted in Algorithm~\ref{alg:eager-ktruss}. From now on we will refer to these as the $s_{12}$ and $S_{22}$ update rules, respectively. As data dependencies exist among different iterations, the updates are performed in an atomic fashion.

In essence, the Eager algorithm in Algorithm \ref{alg:eager-ktruss} computes edge supports by performing set intersections for vertex neighborhoods on either side of a current edge and updating all edges in each triangle found. Neighborhood intersections are grouped by their common source vertex, whose outgoing neighborhood is $a_{12}$. Imbalances can occur since vertices do not have identically sized outgoing neighborhoods.

\begin{algorithm}[tb]
  \footnotesize
  \KwIn{$A$ is the upper-triangular adjacency matrix of the input graph}
  \KwOut{$S$ represents the support of edges in $A$}
  \BlankLine
  \SetKwFor{ParFor}{parallel for}{do}{end}
  \ParFor{ $i\ in\ range(0,numRows(A))$ } {
    partition $A$ and $S$ where $\alpha_{11}$ and $\sigma_{11}$ are scalar corresponding to the diagonal values of $A_{ii}$ and $S_{ii}$ as follows:
    {\scriptsize $\left(\begin{array}{c|cc}
    A_{00} & a_{01} & A_{02} \\ \hline
    \quad  & \alpha_{11} & a_{12}^T \\
    \quad  & \quad       & A_{22} \\
    \end{array}\right)
    $ } and
    {\scriptsize $\left(\begin{array}{c|cc}
    S_{00} & s_{01} & S_{02} \\ \hline
    \quad  & \sigma_{11} & s_{12}^T \\
    \quad  & \quad       & S_{22} \\
    \end{array}\right)
    $ }
    \\
    $s_{12}^T \leftarrow s_{12}^T + a_{12}^T A_{22} \circ a_{12}^T$\\ 
    $S_{22}^T \leftarrow S_{22}^T + a_{12} a_{12}^T \circ A_{22}$\\
  }
  \Return{$S$}\\
  \caption{computeSupports in Eager K-truss. Greek, lower-case and upper-case letters are scalars, vectors and matrices respectively. Loop range is defined with a start value and a non-inclusive upper bound.}
  \label{alg:eager-ktruss}
\end{algorithm}
\section{Fine-grained Parallel Eager K-truss}
\label{sec:fine-grained-eager}
In this section, we discuss the source of load imbalance in the existing coarse-grained approach and describe our fine-grained solution. We also discuss implementation details of our approach using the Kokkos framework.



\subsection{Source of Load Imbalance in Eager K-truss}


For each matrix partitioning $i$ at a given iteration, the $s_{12}$ and $S_{22}$ update rules are applied. 
As matrix $A$ is upper triangular,  the computational cost of the two updates may decrease as the iteration count $i$ increases. 
Further, the imbalanced nature of the graph connectivity can lead to significant performance degradation especially when a large number of computing units are used in parallel. Intuitively, this imbalance arises in part from the decrease in length of vector $a_{i_{12}}^T$ and also from the shrinking size, both in width and height, of the sub-matrix $A_{i_{22}}$. However, because the matrix $A$ is typically highly sparse, the width and height of $A_{i_{22}}$ is far less relevant than the number of non-zero elements in $a_{i_{12}}^T$. Even the number of non-zero elements in $A_{i_{22}}$ does not directly influence imbalance. The major reason for this is that the number of row vectors in $A_{i_{22}}$ that are relevant to the update rules is directly determined by the non-zero elements present in $a_{i_{12}}^T$. 


\subsection{Fine-grained Tasks for Updating Edge Supports}

\begin{algorithm}[tb]
  \footnotesize
  \KwIn{$A$ is the upper-triangular adjacency matrix of the input graph}
  \KwOut{$S$ represents the support of edges in $A$}
  \BlankLine
  \SetKwFor{ParFor}{parallel for}{do}{end}
  \ParFor{ $i\ in\ range(0,numRows(A))$ } {
    partition $A$ and $S$ where $\alpha_{11}$ and $\sigma_{11}$ are scalar corresponding to the diagonal values of $A_{ii}$ and $S_{ii}$ as follows:
    {\scriptsize $\left(\begin{array}{c|cc}
    A_{00} & a_{01} & A_{02} \\ \hline
    \quad  & \alpha_{11} & a_{12}^T \\
    \quad  & \quad       & A_{22} \\
    \end{array}\right)
    $ } and
    {\scriptsize $\left(\begin{array}{c|cc}
    S_{00} & s_{01} & S_{02} \\ \hline
    \quad  & \sigma_{11} & s_{12}^T \\
    \quad  & \quad       & S_{22} \\
    \end{array}\right)
    $ }
    \\
    \ParFor{$j\ in\ nonZeros(a_{12}^T)$} {
      $\kappa \leftarrow a_{12}^T(j)$ \tcp*[h]{index of $A_{22}$ row w.r.t $A$} \\      
      \tcp{$A(\kappa,:)$ is the row in $A_{22}$ for task $(i,j)$} 
      \tcp{$S(\kappa,:)$ is the row in $S_{22}$ for task $(i,j)$}
      $s^T_{12}(j) \leftarrow s^T_{12}(j) + A_{22}(\kappa, :)a_{12}$ \\
      $s_{12}^T(j+1:) \leftarrow s_{12}^T(j+1:) + a_{12}^T(j+1:) \circ A(\kappa,:)$\\
      $S(\kappa,:) \leftarrow S(\kappa,:) + a_{12}^T(j+1:) \circ A(\kappa,:)$\\
    }
  }
  \Return{$S$}
  \caption{computeSupports in fine-grained Eager K-truss. 
  Greek, lower-case and upper-case letters are scalars, vectors and matrices respectively.
  Loop and vector or matrix dimension ranges are defined with a start value, colon, and either a non-inclusive upper bound or no upper limit indicating inclusion of the remaining space.}
  \label{alg:fine-ktruss}
\end{algorithm}

\begin{figure}[tb]
    \centering
    \includegraphics[width=\linewidth]{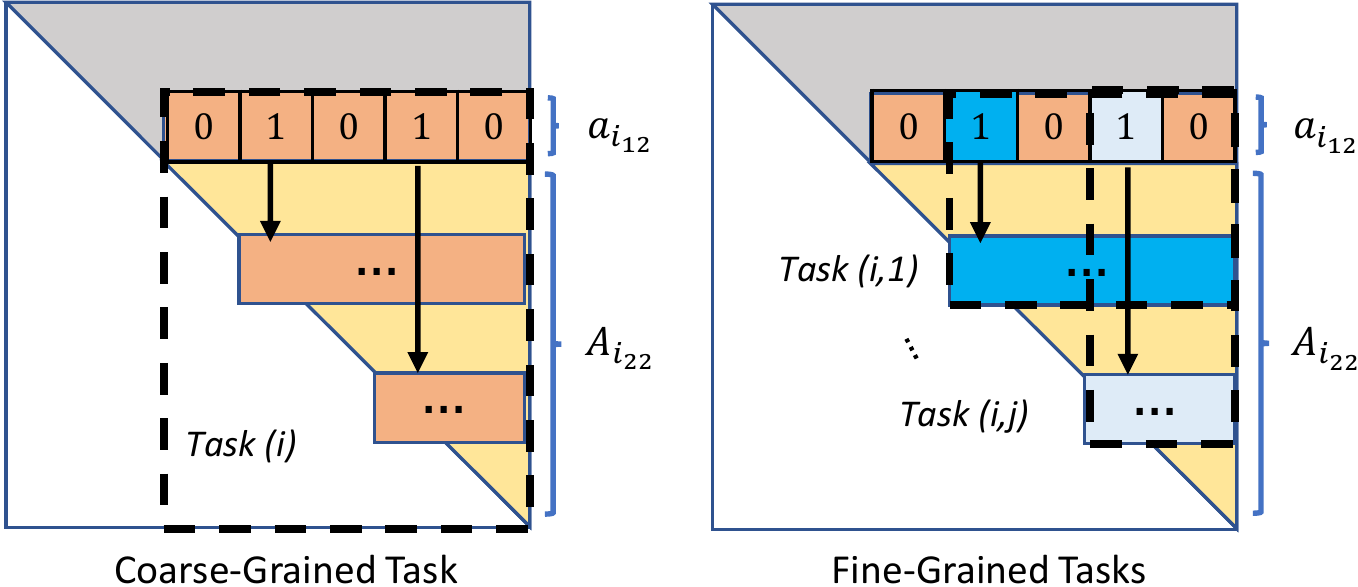}
    \caption{Partitioning $i$ corresponding to a specific set of inputs to the support updates. The left side illustrates that the 
    the work is roughly proportional to the number of non-zero elements in $a_{i_{12}}^T$. The right side shows a fine-grained set of tasks, where each task is further refined by the $i^{th}$ partitioning and the $j^{th}$ non-zero value in $a_{i_{12}}^T$. Task $(i,j)$ updates $s_{12}$  on the $i^{th}$ row, and a row in $S_{22}$ indicated by the $j^{th}$ value in $a_{i_{12}}^T$.}
    \label{fig:diff_tasks}
    \vspace{-4mm}
\end{figure}

Our load balancing strategy is to take advantage solely of the information provided by a particular $a_{i_{12}}^T$: the number of non-zero elements which directly indicate the number of neighborhoods in $A_{i_{22}}$ that must be visited. Thus, a fine-grained task can be defined as a row-wise update by selecting a row in $A_{i_{22}}$ based on the $j^{th}$ value of $a_{i_{12}}^T$, rather than updating multiple rows in $A_{i_{22}}$ from the outer product of $a_{i_{12}}^T$. As a result, our fine-grained Eager algorithm iterates over a pair iterator $(i,j)$ that corresponds to the number of non-zero elements of $A$. Fig.~\ref{fig:diff_tasks} illustrates the difference between the two parallel approaches. To distinguish these two algorithms, we refer the original Eager K-truss algorithm as \textit{coarse-grained} parallel since each parallel task in the algorithm operates on a group of edges that share the same source vertex. In contrast, we refer to our parallel approach using tasks defined by each non-zero element of the matrix as \textit{fine-grained} parallel; it is described in Algorithm~\ref{alg:fine-ktruss}.

The actual work to be performed 
is determined by both the number of non-zero elements in $a_{i_{12}}^T$ \textit{and} 
the non-zero elements within the relevant row of $A_{i_{22}}$ indicated by a non-zero value in $a_{i_{12}}^T$. Therefore, the update rules could then further divided into smaller tasks by grouping the elements within a row of $A_{i_{22}}$. However the information as to exactly how many elements of a row of $A_{i_{22}}$ should be grouped is not straightforward and precise determination of this information can cost similarly to the actual cost of applying the update rules. Therefore we leave it to future work to explore the merits of such ultra-fine-grained tasks. Finally, we note that in either approach outlined above, the Eager K-truss algorithm is still edge-centric; it is simply that the expression of parallelism is changed from parallel over groups of edges within a row (on the order of number of vertices) to parallel over tasks identified by individual edges (on the order of non-zero elements).

\begin{minipage}{0.97\linewidth}
\begin{lstlisting}[language=C++, caption=Fine-grained support computation in  Kokkos,label=lst:fine-grained-eager-kokkos]
using namespace Kokkos;
using SpT = Cuda; // Serial, OpenMP, Cuda
using IJ_ViewType = View<uint32_t*,SpT>;
// ++, += operators are overloaded by atomic updates
using S_ViewType = View<uint16_t,SpT,MemoryTraits<Atomic> >;
// CSR input (IA and JA) and support matrix S are stored on device memory
uint32_t fine_grained_parallel_eager(
  IJ_ViewType IA, IJ_ViewType JA, // CSR input of a graph (matrix A) 
  S_ViewType S, // support of edges
  uint32_t K) // K-truss parameter
  const uint32_t nnz = JA.extent(0), nverts = IA.extent(0)-1;
  uint32_t tri_new(1);
  while (tri_new) {
    tri_new = 0;
    // Step 1: computeSupports
    parallel_for(RangePolicy<SpT>(0, nnz),
      KOKKOS_LAMBDA(const uint32_t ij) const {
        auto a12_pred = JA(i);
        if (a12_pred != 0) {
          auto a12_start_off = i+1;
          auto A22_start_off = IA(a12_pred);
          auto A22_end_off = IA(a12_pred+1)-1;
          uint16_t SL(0);
          while (JA(a12_start_off)!=0 && JA(A22_start_off)!=0){
            const auto flag = (JA(A22_start_off) == JA(a12_start_off));
            if (flag) {
              ++(S(a12_start_off));
              ++(S(A22_start_off));
              SL++;
              a12_start_off++;
              A22_start_off++;
            } else if (JA(A22_start_off) > JA(a12_start_off)){
              a12_start_off++;
            } else {
              A22_start_off++;
            }
          }
          S(i) += SL;
        }
      });
    fence();
    pruneEdges();   // Step 2 (omitted due to space constraints)
  }
  return tri_new;
}
\end{lstlisting}
\vspace{-3mm}
\end{minipage}


\subsection{Performance Portable Implementation via Kokkos}

Kokkos provides a performance portable parallel programming model with node-level device abstractions i.e., execution space, memory space, execution policy and parallel patterns. 
We briefly explain Kokkos features that we primarily used for our fine-grained-parallel implementation. An execution space and a memory space define where the code is executed and where data resides. For example, an execution space supports Serial, OpenMP and Cuda for GPUs while a memory space includes HostSpace, CudaUVMSpace (host accessible) and CudaSpace. Then, a device can be described as a combination of an execution space and corresponding memory space. An execution policy defines how computing units are mapped to concurrent tasks. In our implementation, we use  RangePolicy that maps a single thread to a single work unit defined in a range of an iteration space. Kokkos also provides commonly used parallel algorithms i.e., parallel for, reduce, and scan, which are interfaced via a functor provided by users. 

Listing~\ref{lst:fine-grained-eager-kokkos} illustrates the implementation of fine-grained support computation using Kokkos. The pruning subroutine is identical to \cite{low2018linear} and is omitted for brevity. As input, the code expects an upper-trianglular CSR input matrix of $IA$ and $JA$ arrays where they represent row pointers and column indices. $S$ indicates an array for storing edge supports and the array is decorated with an Atomic memory trait. Using Kokkos, this code works for both CPUs and NVIDIA GPUs by changing a single line of the code specifying an execution space. Note that while we use a flat range policy in the implementation, Algorithm~\ref{alg:fine-ktruss} is expressed as nested parallel for readability. 

\subsection{Zero-terminated CSR}
Recognize that  
the information determining which $a_{i,j_{12}}$ and $A_{i,j_{22}}$ (notated $A(\kappa, :)$ in Algorithm~\ref{alg:fine-ktruss}) sub-vectors are used as inputs to a particular task is almost entirely characterized by the location and contents of the element in $a_{i,j_{12}}$ located at $A(i,j)$. By defining each task based on the matrix element, we are able to avoid adding an additional data structure for tracking tasks. 

However, we did modify the triangularized CSR graph representation such that 
each vertex neighborhood is terminated with an additional zero value. This increases the length of the entire vector of non-zero elements by the number of vertices, but allows the ends of each task's input vectors to be implicitly coded without external bookkeeping. 
This change is minor because the pruning step already introduces zeros as a way of implementing early termination. Therefore, zero-terminating the rows at the start simplifies the implementation and enables fine-grained parallel implementation.

\section{Experimental Results and Analysis}
In this section we cover our experimental setup for testing 
performance of our parallel implementations 
on CPU and GPU. We then present and discuss selected performance results
on a set of graphs.
\begin{figure*}[ht]
    \begin{center}
        \includegraphics[height=6cm]{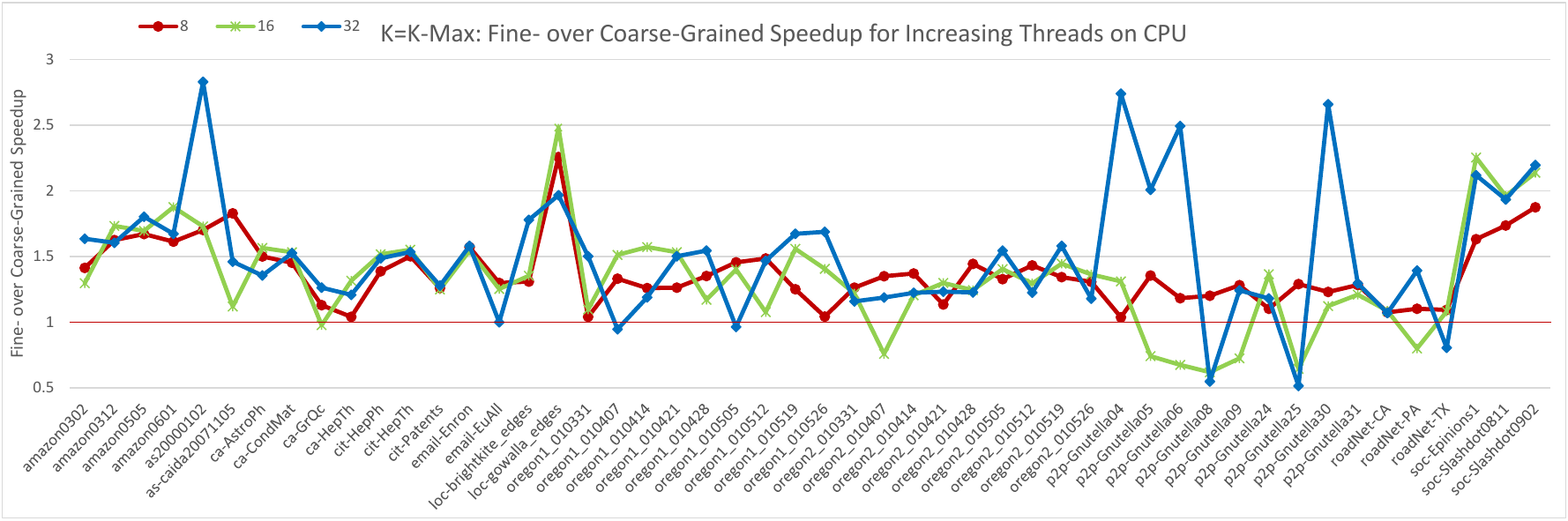}
        \label{fig:Kmax_thread_scaling}
    \end{center}
\vspace{-4mm}
\caption{Speedup of fine- over coarse-grained performance for $K= K_{max}$ for each graph and number of threads. Speedup improvement from threading is highly graph-dependent. The red line indicates coarse-grained the normalization baseline.}
\label{fig:thread_scaling}
\vspace{-4mm}
\end{figure*}

\subsection{Experimental Setup}
Input graphs from the Stanford Network Analysis Project (SNAP) \cite{snapnets} were downloaded from the GraphChallenge collection~\cite{graphchallenge:Isomorphism:2017}.
These graphs were made upper-triangular before being used as inputs.
Our CPU test platform was a single-node, dual-socket machine with 
two 24-core, 48-thread Intel Xeon Platinum 8160 CPUs, each operating at
2.10 GHz with a total of 187 GB of main memory.
Our GPU is an Nvidia Tesla V100 hosted on a second machine with a Power9 IBM processor. 
Experiments were run for the cases when $K=3$, and $K=K_{max}$, i.e. the largest value of $K$ that returns a non-empty K-truss. 
The $K=3$ case is consistent with experiments in Bisson et al~\cite{Bisson2017, Bisson2018}.
Performance numbers are provided in terms of millions of edges processed per second 
(ME/s), and the mean of 10 trials is reported.
Correctness of all implementations was verified against the
reference K-truss code provided by Low et al.~\cite{low2018linear}.



Table~\ref{tab:48T_K3} reports runtimes and ME/s results comparing between the coarse- and fine-grained using 48 CPU threads and on the GPU for the $K=3$ case. Due to space constraints, we do not also include a similar table for $K= K_{max}$. Where named in a plot, graphs are ordered from least number of edges (non-zero entries) to greatest, left to right. Where a plot bar for coarse-grained performance is not visible, vertical text is placed to state the ME/s rate achieved.

\subsection*{CPU Performance}

Fig. \ref{fig:thread_scaling} shows speedup of fine-grained versus coarse-grained parallelism for $K= K_{max}$. We omit the plot for $K=3$ except to state that speedups are slightly higher.

\begin{figure*}
    \centering
    \includegraphics[height=1.75in,width=\textwidth]{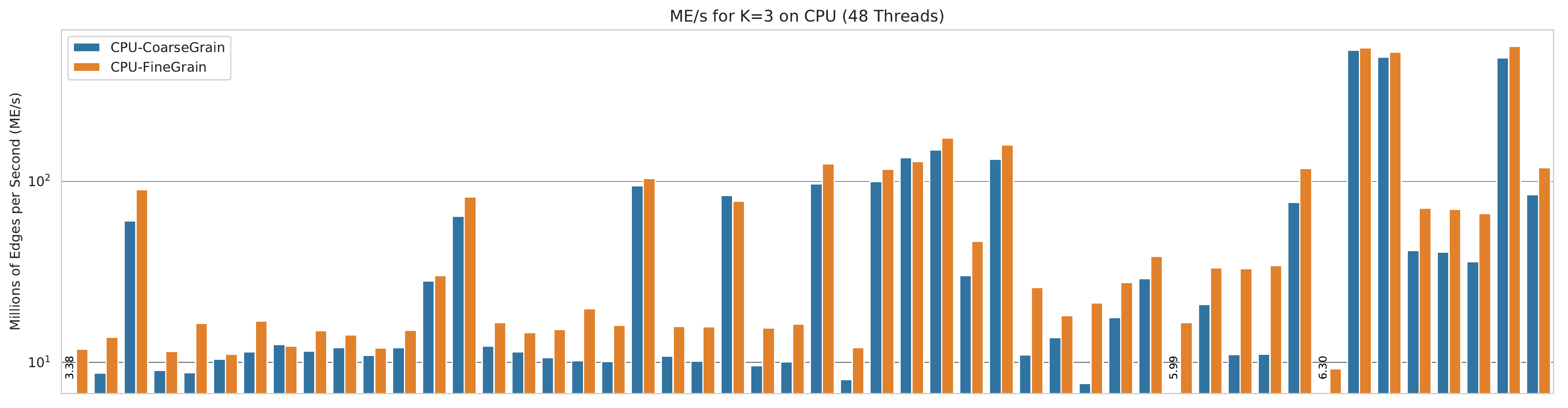}
    \includegraphics[height=2.5in,width=\textwidth]{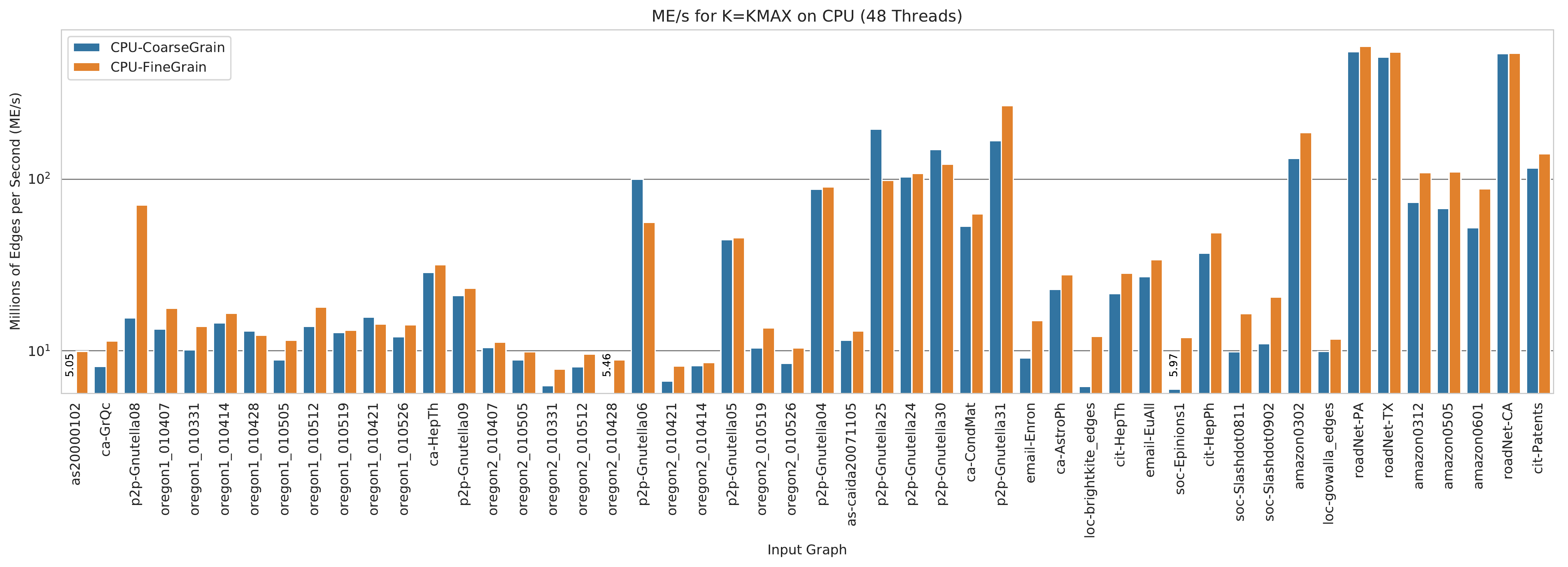}
    \caption{Performance for 48 CPU threads of coarse- and fine-grained implementations. Top: $K=3$. Bottom: $K=K_{max}$.}
    \label{fig:CPU48T_all_graphs}
\end{figure*}

Speedup is above unity for most graphs, indicating that fine-grained has better performance per graph and per number of threads.
Some larger graphs show drops in speedup below unity for threads greater than 8.
This may be because $K=K_{max}$ aggressively prunes edges from the graph and reduces overall opportunities for parallelism.
Some graphs show peaks in speedup for 32 threads even if the 16 thread run has worse speedup than 8.
In future work we will examine the graphs' properties to determine causes for these peaks and troughs.
Fig. \ref{fig:CPU48T_all_graphs} shows performance of the CPU parallel implementations across all input graphs for 48 threads. 
Overall, the geometric mean speedup across graphs of the fine-grained parallel implementation over coarse-grained is 1.48x for $K=3$ and 1.26x for $K= K_{max}$, both for 48 threads.



\begin{figure*}
    \centering
    \includegraphics[height=1.75in,width=\textwidth]{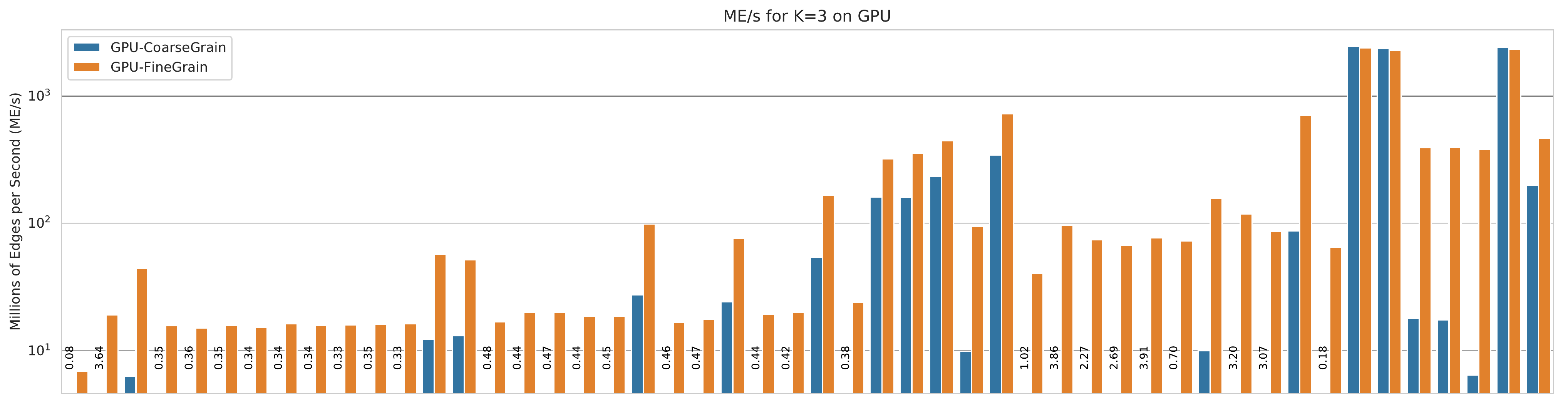}
    \includegraphics[height=2.5in,width=\textwidth]{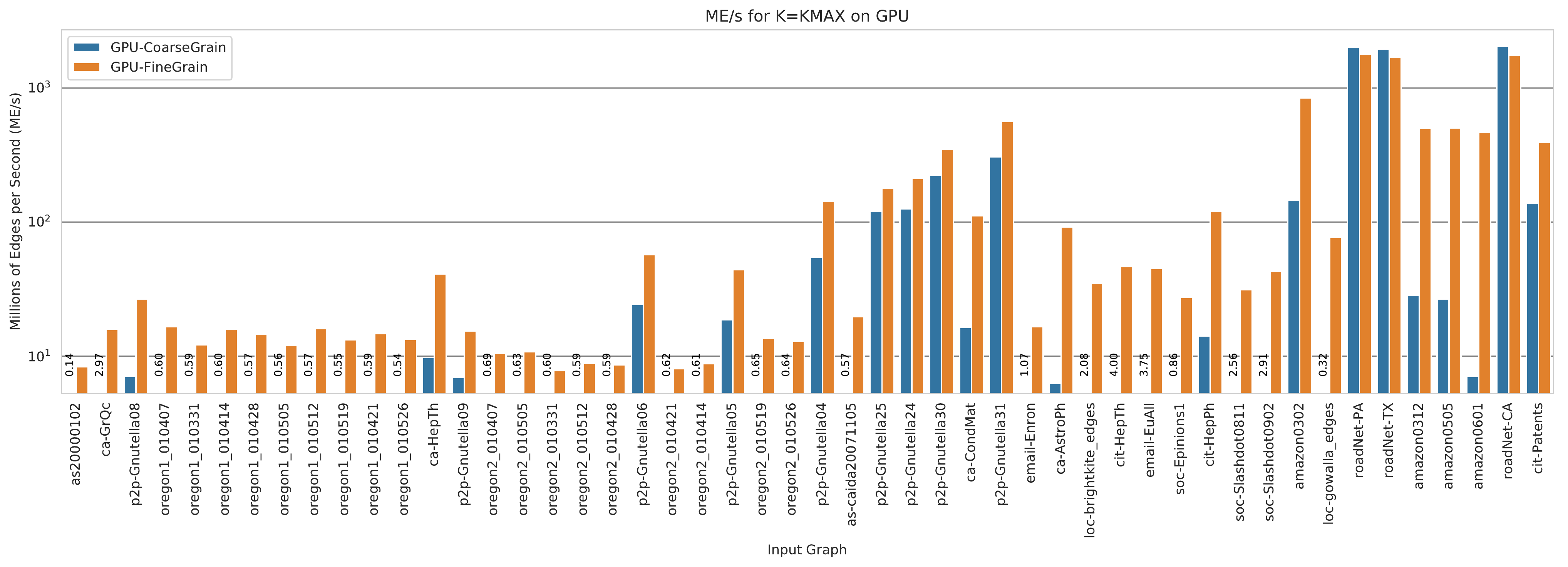}
    \caption{Performance on GPU of coarse- and fine-grained implementations. Top: $K=3$. Bottom: $K=K_{max}$.}
    \label{fig:GPU_all_graphs}
    \vspace{-2mm}
\end{figure*}


\subsection*{GPU Performance}
The difference in performance between the two parallelism approaches is much larger on the GPU.
Fig. \ref{fig:GPU_all_graphs} shows that, for the majority of the input graphs, the fine-grained implementation achieves significantly higher performance. For the case of $K=3$, the mean geometric speedup between the two parallel schemes on the GPU is 16.93x; for $K=K_{max}$, it is 9.97x.

In addition to reducing load imbalance, executing in parallel over fine-grained tasks appears to benefit the GPU by increasing the amount of parallel work. This seems to be the case for the \textit{oregon} networks and \textit{as-caida20071105}, which have on the order 10,000 - 1,000 vertices and are among the smallest. 
However, larger graphs such as the \textit{roadNet} graphs and \textit{cit-Patents} show lower performance on the GPU. 
Determining the sources of these performance degradations will be the subject of future work. 
Overall, the fine-grained GPU implementation are 1.92x and 1.56x faster, for $K=3$ and $K_{max}$ respectively, than the fine-grained CPU implementation.

\begin{table*}[]
\scriptsize
\centering
\caption{Runtimes and ME/s Performance Numbers for all graphs tested. Postfix in subgroup heading refers to coarse or fine-grained parallelism. All CPU results presented in this table are for 48 threads, and all runs are for $K=3$.}
\resizebox{\textwidth}{!}{
\begin{tabular}{|l|cc|cccc|cccc|}
\hline
 & & &  \multicolumn{4}{c|}{Time (ms)} & \multicolumn{4}{c|}{Millions of Edges per Second (ME/s)} \\
 Input Graph           & Vertices (k) & Edges    & CPU-C     & CPU-F   & GPU-C    & GPU-F  & CPU-C                        & CPU-F   & GPU-C    & GPU-F    \\ \hline
ca-GrQc               & 5.2k         & 14.5k    & 1.660     & 1.051   & 3.982    & 0.762  & 8.724                        & 13.784  & 3.637    & 19.003   \\
p2p-Gnutella08        & 6.3k         & 20.8k    & 0.343     & 0.230   & 3.334    & 0.472  & 60.663                       & 90.178  & 6.232    & 44.028   \\
as20000102            & 6.5k         & 12.6k    & 3.715     & 1.062   & 148.729  & 1.837  & 3.384                        & 11.839  & 0.085    & 6.843    \\
p2p-Gnutella09        & 8.1k         & 26.0k    & 0.404     & 0.316   & 2.000    & 0.506  & 64.309                       & 82.242  & 13.006   & 51.409   \\
p2p-Gnutella06        & 8.7k         & 31.5k    & 0.333     & 0.303   & 1.153    & 0.320  & 94.727                       & 104.112 & 27.342   & 98.516   \\
p2p-Gnutella05        & 8.8k         & 31.8k    & 0.380     & 0.409   & 1.326    & 0.417  & 83.831                       & 77.808  & 24.015   & 76.316   \\
ca-HepTh              & 9.9k         & 26.0k    & 0.924     & 0.860   & 2.135    & 0.458  & 28.115                       & 30.191  & 12.164   & 56.660   \\
oregon1\_010331       & 10.7k        & 22.0k    & 2.511     & 1.338   & 61.248   & 1.475  & 8.763                        & 16.448  & 0.359    & 14.918   \\
oregon1\_010407       & 10.7k        & 22.0k    & 2.433     & 1.916   & 62.416   & 1.408  & 9.040                        & 11.484  & 0.352    & 15.628   \\
oregon1\_010414       & 10.8k        & 22.5k    & 2.161     & 2.023   & 63.569   & 1.428  & 10.396                       & 11.106  & 0.353    & 15.730   \\
oregon1\_010421       & 10.9k        & 22.7k    & 2.081     & 1.892   & 64.603   & 1.421  & 10.932                       & 12.021  & 0.352    & 16.011   \\
p2p-Gnutella04        & 10.9k        & 40.0k    & 0.413     & 0.319   & 0.740    & 0.241  & 96.838                       & 125.216 & 54.024   & 166.088  \\
oregon1\_010428       & 10.9k        & 22.5k    & 1.964     & 1.330   & 66.396   & 1.482  & 11.453                       & 16.916  & 0.339    & 15.174   \\
oregon2\_010331       & 10.9k        & 31.2k    & 2.938     & 2.049   & 65.880   & 1.568  & 10.613                       & 15.216  & 0.473    & 19.881   \\
oregon1\_010505       & 10.9k        & 22.6k    & 1.801     & 1.842   & 66.031   & 1.399  & 12.552                       & 12.272  & 0.342    & 16.163   \\
oregon2\_010407       & 11.0k        & 30.9k    & 2.515     & 1.860   & 64.638   & 1.846  & 12.269                       & 16.588  & 0.477    & 16.715   \\
oregon1\_010512       & 11.0k        & 22.7k    & 1.961     & 1.518   & 66.446   & 1.443  & 11.562                       & 14.937  & 0.341    & 15.713   \\
oregon2\_010414       & 11.0k        & 31.8k    & 3.120     & 2.020   & 67.370   & 1.816  & 10.180                       & 15.727  & 0.471    & 17.488   \\
oregon1\_010519       & 11.1k        & 22.7k    & 1.882     & 1.600   & 68.218   & 1.438  & 12.076                       & 14.199  & 0.333    & 15.800   \\
oregon2\_010421       & 11.1k        & 31.5k    & 2.917     & 2.002   & 68.057   & 1.899  & 10.813                       & 15.756  & 0.463    & 16.610   \\
oregon2\_010428       & 11.1k        & 31.4k    & 3.107     & 1.960   & 70.229   & 1.710  & 10.116                       & 16.035  & 0.448    & 18.380   \\
oregon2\_010505       & 11.2k        & 30.9k    & 2.703     & 2.122   & 70.168   & 1.550  & 11.447                       & 14.583  & 0.441    & 19.967   \\
oregon1\_010526       & 11.2k        & 23.4k    & 1.945     & 1.554   & 70.848   & 1.445  & 12.034                       & 15.067  & 0.330    & 16.206   \\
oregon2\_010512       & 11.3k        & 31.3k    & 3.060     & 1.585   & 70.707   & 1.687  & 10.229                       & 19.753  & 0.443    & 18.551   \\
oregon2\_010519       & 11.4k        & 32.3k    & 3.372     & 2.085   & 74.135   & 1.696  & 9.575                        & 15.482  & 0.436    & 19.041   \\
oregon2\_010526       & 11.5k        & 32.7k    & 3.253     & 2.011   & 77.051   & 1.639  & 10.061                       & 16.274  & 0.425    & 19.976   \\
ca-AstroPh            & 18.8k        & 198.1k   & 14.461    & 10.928  & 51.303   & 2.055  & 13.695                       & 18.123  & 3.860    & 96.365   \\
p2p-Gnutella25        & 22.7k        & 54.7k    & 0.548     & 0.468   & 0.340    & 0.171  & 99.790                       & 116.791 & 160.755  & 320.662  \\
ca-CondMat            & 23.1k        & 93.4k    & 3.090     & 1.996   & 9.496    & 0.990  & 30.239                       & 46.804  & 9.840    & 94.431   \\
as-caida20071105      & 26.5k        & 53.4k    & 6.659     & 4.417   & 139.697  & 2.238  & 8.016                        & 12.085  & 0.382    & 23.847   \\
p2p-Gnutella24        & 26.5k        & 65.4k    & 0.483     & 0.507   & 0.410    & 0.186  & 135.452                      & 129.009 & 159.475  & 352.204  \\
cit-HepTh             & 27.8k        & 352.3k   & 19.929    & 12.755  & 131.030  & 5.291  & 17.677                       & 27.619  & 2.689    & 66.586   \\
cit-HepPh             & 34.5k        & 420.9k   & 20.176    & 12.628  & 42.338   & 2.693  & 20.860                       & 33.328  & 9.941    & 156.291  \\
p2p-Gnutella30        & 36.7k        & 88.3k    & 0.593     & 0.507   & 0.381    & 0.198  & 148.951                      & 174.183 & 231.832  & 446.326  \\
email-Enron           & 36.7k        & 183.8k   & 16.768    & 7.101   & 180.731  & 4.599  & 10.963                       & 25.887  & 1.017    & 39.975   \\
loc-brightkite\_edges & 58.2k        & 214.1k   & 28.003    & 10.038  & 94.141   & 2.903  & 7.645                        & 21.326  & 2.274    & 73.749   \\
p2p-Gnutella31        & 62.6k        & 147.9k   & 1.116     & 0.930   & 0.431    & 0.203  & 132.508                      & 159.058 & 343.376  & 727.099  \\
soc-Epinions1         & 75.9k        & 405.7k   & 67.730    & 24.453  & 582.784  & 5.599  & 5.991                        & 16.593  & 0.696    & 72.472   \\
soc-Slashdot0811      & 77.4k        & 469.2k   & 42.498    & 14.202  & 146.617  & 3.968  & 11.040                       & 33.037  & 3.200    & 118.232  \\
soc-Slashdot0902      & 82.2k        & 504.2k   & 45.469    & 14.729  & 164.038  & 5.865  & 11.090                       & 34.233  & 3.074    & 85.977   \\
loc-gowalla\_edges    & 196.6k       & 950.3k   & 150.897   & 103.023 & 5332.719 & 14.762 & 6.298                        & 9.224   & 0.178    & 64.376   \\
amazon0302            & 262.1k       & 899.8k   & 11.741    & 7.625   & 10.346   & 1.275  & 76.634                       & 118.009 & 86.967   & 705.830  \\
email-EuAll           & 265.0k       & 364.5k   & 12.535    & 9.439   & 93.244   & 4.771  & 29.078                       & 38.616  & 3.909    & 76.389   \\
amazon0312            & 400.7k       & 2349.9k  & 56.524    & 33.074  & 131.514  & 5.975  & 41.573                       & 71.049  & 17.868   & 393.303  \\
amazon0601            & 403.4k       & 2443.4k  & 67.959    & 36.734  & 383.056  & 6.454  & 35.954                       & 66.516  & 6.379    & 378.594  \\
amazon0505            & 410.2k       & 2439.4k  & 60.062    & 34.748  & 140.891  & 6.161  & 40.615                       & 70.204  & 17.314   & 395.923  \\
roadNet-PA            & 1088.1k      & 1541.9k  & 2.894     & 2.821   & 0.627    & 0.644  & 532.736                      & 546.617 & 2458.775 & 2395.740 \\
roadNet-TX            & 1379.9k      & 1921.7k  & 3.955     & 3.696   & 0.812    & 0.837  & 485.881                      & 519.887 & 2367.159 & 2296.164 \\
roadNet-CA            & 1965.2k      & 2766.6k  & 5.733     & 4.956   & 1.149    & 1.189  & 482.601                      & 558.234 & 2407.210 & 2326.053 \\
cit-Patents           & 3774.8k      & 16518.9k & 195.765   & 138.447 & 82.991   & 35.532 & 84.382                       & 119.316 & 199.046  & 464.903  \\
\hline
\end{tabular}
}
\label{tab:48T_K3}
\end{table*}



\section{Conclusions and Future Work}
This paper presents a Kokkos implementation of the Eager K-truss algorithm 
and improves performance by exploiting fine-grained parallelism. 
We presented results for our fine-grained approach of Eager K-truss on both CPU and GPU, showing that fine-grained parallelism gives significantly improved performance on GPUs and some benefits on the CPU system. By switching parallelism strategies, we observe a mean improvement of 16.93x and 9.97x on the GPU for $K=3$ and $K=K_{max}$, respectively, and an improvement on the CPU of 1.48x and 1.26x for the two $K$ settings.


An alternative to the presented parallel approach is hierarchical parallelism (team parallelism in Kokkos) for expression of ultra-fine-grained parallelism and further improved workload balancing. In addition, while the fine-grained approach generally performs well on both platforms, larger graphs tend to show nearly the same performance on the GPU for both approaches. Determining the source of this behavior will be the subject of future work.

Graph rebalancing via full or partial sorting is another strategy to improve load-balancing \cite{Balaji:2019:CDD:3307681.3326609}. This approach could complement our fine-grained parallelism for Eager K-truss and warrants further investigation. Finally, our fine-grained approach may enable speedups in other motif-counting problems, such as butterfly counting for bipartite networks \cite{Wang2014,sanei-mehri_butterfly_2018}.

\section*{Acknowledgments}

Sandia National Laboratories is a multimission laboratory managed and operated by National Technology and Engineering Solutions of Sandia, LLC., a wholly owned subsidiary of Honeywell International, Inc., for the U.S. Department of Energy's National Nuclear Security Administration under contract DE-NA-0003525.

This material is based upon work funded and supported by the Department of Defense under Contract No. FA8702-15-D-0002 with Carnegie Mellon University for the operation of the Software Engineering Institute, a federally funded research and development center.
DM19-0868.

\bibliographystyle{IEEEtran}
\bibliography{references}

\end{document}